\documentclass[aps,pre,reprint,nofootinbib,superscriptaddress,showpacs]{revtex4-1}

\usepackage{graphicx}
\usepackage{epsfig}
\usepackage{dcolumn}
\usepackage{bm}
\usepackage{hyperref}
\hypersetup{
colorlinks,
citecolor=blue,
filecolor=blue,
linkcolor=blue,
urlcolor=blue,
pdftitle=title,
pdfauthor=author,
bookmarks,}
\usepackage{amsmath}
\usepackage{amssymb}
\usepackage{longtable}
\begin{document}


\title
{Stratification, segregation and mixing of granular materials in quasi-2D bounded heaps}
\author{Yi Fan}
\affiliation{Department of Mechanical Engineering, Northwestern University, Evanston, Illinois 60208, USA}

\author{Youcef Boukerkour}
\author{Thibault Blanc}
\affiliation{French Air Force Academy, Salon de Provence, France}

\author{Paul B. Umbanhowar}
\affiliation{Department of Mechanical Engineering, Northwestern University, Evanston, Illinois 60208, USA}

\author{Julio M. Ottino}
\affiliation{Department of Mechanical Engineering, Northwestern University, Evanston, Illinois 60208, USA}
\affiliation{Department of Chemical and Biological Engineering, Northwestern University, Evanston, Illinois 60208, USA}
\affiliation{The Northwestern University Institute on Complex Systems (NICO), Northwestern University, Evanston, Illinois 60208, USA}

\author{Richard M. Lueptow}
\email{r-lueptow@northwestern.edu}
\affiliation{Department of Mechanical Engineering, Northwestern University, Evanston, Illinois 60208, USA}

\date{\today}

\begin{abstract}
Segregation and mixing of granular mixtures during heap formation have important consequences in industry and agriculture. This research investigates three different final particle configurations of bi-disperse granular mixtures - stratified, segregated and mixed - during filling of quasi-two dimensional silos. We consider a larger number and relatively wider range of control parameters than previous studies, including particle size ratio, flow rate, system size and heap rise velocity. The boundary between stratified and unstratified states is primarily controlled by the two-dimensional flow rate, with the critical flow rate for the transition depending weakly on particle size ratio and flowing layer length. In contrast, the transition from segregated to mixed states is controlled by the rise velocity of the heap, a control parameter not previously considered. The critical rise velocity for the transition depends strongly on the particle size ratio.


\end{abstract}


\maketitle

\section{Introduction}
\label{Intro}

Heap flow of granular materials occurs in many contexts \cite{Evesque,deGennes}. For example, when granular materials such as powders, grains, or pelletized polymers flow into the top of a container, a heap builds, where the granular material tumbles down the pile via a flowing layer that is a few particle diameters thick. When materials are mixtures of particles differing in size, density, shape, and/or surface properties, different components tend to distribute inhomogeneously. In some cases, larger particles flow further down the heap than smaller particles resulting in segregation \cite{Williams1963,Williams1968,Drahun,Shinohara1972,Shinohara1990,Thomas2,Goyal,Rahman2011}. In other cases, large and small particles form alternating layers resulting in stratification \cite{Makse1997,MaksePRL,Grasselli,Koeppe,Baxter,Shimokawa2007,Shimokawa2008}. In still other situations, large and small particles remain mixed \cite{Baxter}.

In what were perhaps the earliest attempts to understand the physical mechanisms driving segregation in heap flow, Williams \cite{Williams1963,Williams1968} and Drahun and Bridgwater \cite{Drahun} performed heap flow experiments using bi-disperse mixtures of different-sized spherical particles. They proposed a percolation mechanism for heap segregation, in which small particles tend to sink through voids preferentially, while large particles rise to the free surface and roll to the end of the flowing layer. Consequently, small particles accumulate below the upstream portion of the flowing layer at the center of the heap while large particles accumulate at the downstream end of the flowing layer of the heap adjacent to the bounding outer walls. Based on this picture, Shinohara and co-workers \cite{Shinohara1972,Shinohara1990} developed a screening layer model based on conservation equations incorporating the percolation mechanism. A recent experimental and computational study \cite{Rahman2011} performed to test the screening layer model \cite{Shinohara1990} shows the model captures some key features of heap segregation in certain ranges of experimental parameters. However, some variables in the screening layer model (such as the penetration rate of segregating components and the velocity ratio of different sub-layers) can only be determined by fitting experimental or simulation data, limiting the applicability of this model.

\begin{figure}[]
\includegraphics[width=3.375in]{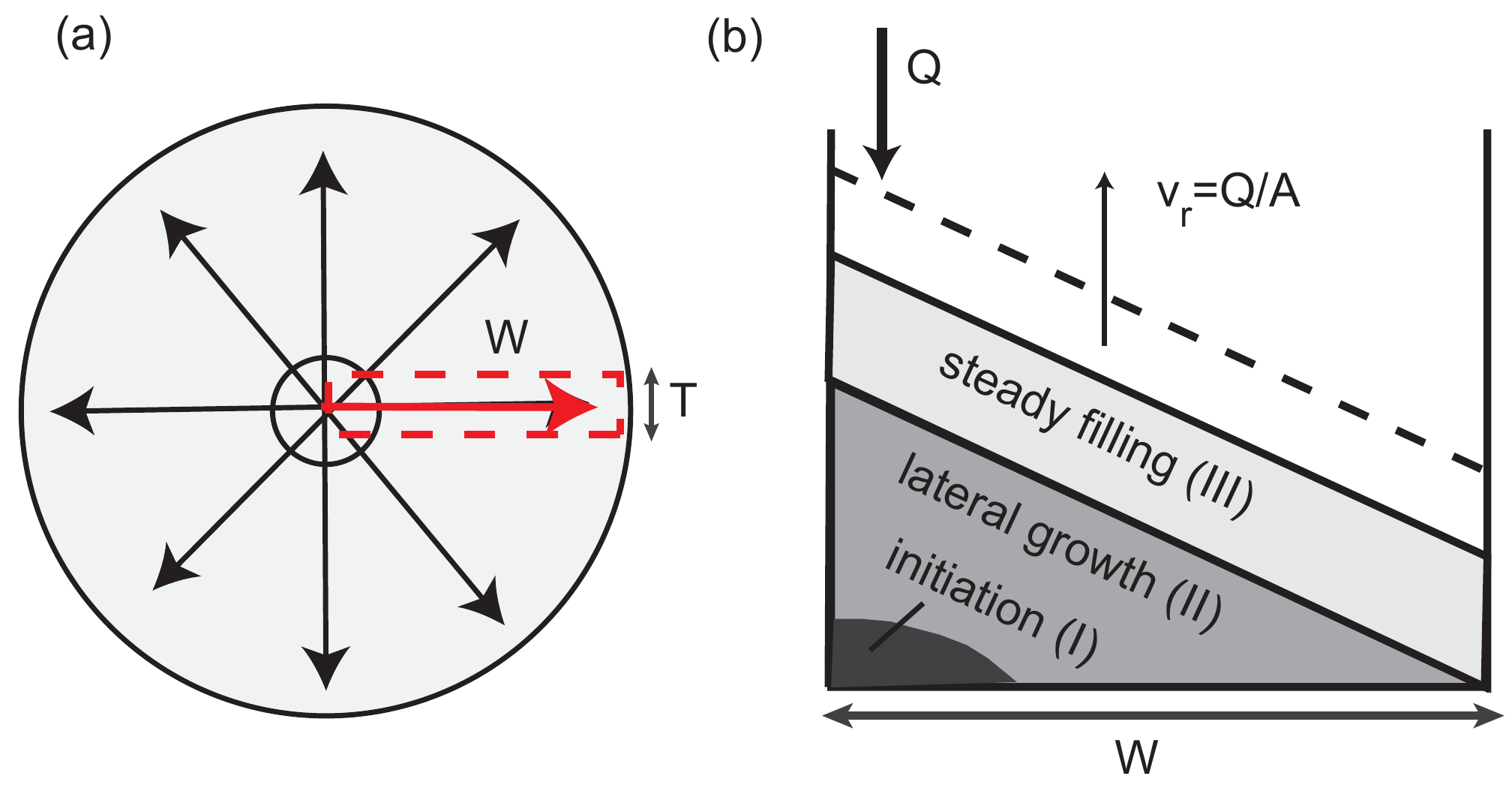}
\caption{\label{stages} (a) Sketch of the top view of a 3D silo, where arrows indicate flow of material. Dashed box shows top view of the quasi-2D silo used here; (b) Side view sketch of a quasi-2D silo rising at the rise velocity $v_r=Q/A$, where $Q$ is volumetric feed rate and $A=TW$. The three stages of heap flow are: (I) Heap initiation; (II) Heap lateral growth; (III) Heap steady filling once it is constrained by the outer wall.}
\end{figure}

Stratification in heap flow was first studied in detail by Makse et al.\ \cite{Makse1997,MaksePRL}\footnote{One figure of Williams \cite{Williams1968} shows stratification in heap flow, but the stratification was not noted.}. In experiments, they found that when the components of granular mixtures differ in both size and shape, the mixtures stratify into alternating layers of larger rough particles and smaller smooth particles. They attributed the formation of stratified layers to competition between size segregation and shape segregation, and proposed that stratification only occurs for mixtures of large rough and small round particles. They also adapted a continuum model \cite{Boutreux,MaksePRL,Cizeau} and a cellular automaton model
\cite{Makse1997,MaksePRL,Cizeau} to successfully reproduce the stratification observed in their experiments. However, Baxter et al.\ \cite{Baxter} found that stratification can also occur for different-sized \textit{smooth spherical} particles, though they did not perform a systematic study over a wide range of parameters. Further, Baxter et al.\ \cite{Baxter} mentioned that a mixed state (no segregation or stratification) exists in certain situations such as at high feed flow rates.
\begin{center}
\begin{table*}
\caption{\label{comparison} Experimental parameters in past and current research}
\begin{ruledtabular}
\begin{tabular}{lcccccc}
References & Segregation type & Size ratio & Silo width &  2D Silo thickness & Flow rate & Heap stage \\
& &  $R$ & $W$ (cm) &  $T$ (cm) &  $Q$ (cm$^3$/s) &  (see Fig.\ \ref{stages})\\
\hline
Williams \cite{Williams1963} & size & 5.2 & 31\footnotemark[1] & 2.5 & unspecified& II \\
Drahun $\&$ Bridgwater \cite{Drahun} & size & 1.3-2.0 & 43 & unspecified & unspecified & III\\
Shinohara et al.\ \cite{Shinohara1972} & size & 14.3 & 30 & unspecified & 30-80 & II$\&$III\\
Shinohara et al.\ \cite{Shinohara1990} & size & 2.0-15.0 & 15-18 & unspecified & 14-33\footnotemark[2] & III\\
Baxter et al.\ \cite{Baxter} & size & 2.0 & 50\footnotemark[1]  & 9 & 6, 736\footnotemark[3]& II\\
Thomas \cite{Thomas2} & size & 2.0-50.0 & 10 & 3D silo & 27-40\footnotemark[1]& II\\
Goyal  $\&$ Tomassone \cite{Goyal} & size & 1.3-5.0 & 22 & 0.5 & unspecified & II\\
Rahman et al.\ \cite{Rahman2011} & size & 9.2-15.2 & 15-18 & 3D silo& 33\footnotemark[2]& III\\
Makse et al.\ \cite{Makse1997,MaksePRL} &  size $\&$ shape & 1.7-6.7 & 30 & 0.5-1 & unspecified & II\\
Grasselli $\&$ Herrmann \cite{Grasselli} &  size $\&$ shape & 1.2-10.5 & 30 & 0.1-0.6 & 0.2-3.5 \footnotemark[3]& II\\
Koeppe et al.\ \cite{Koeppe} &  size $\&$ shape & 2.0 & 27 & 0.3-2.4 & 0.5-7.4\footnotemark[3] & II\\
Shimokawa $\&$ Ohta \cite{Shimokawa2007,Shimokawa2008} &  size $\&$ shape & 2.0-8.0 & 60 & 0.5 & 0.08-1.72\footnotemark[4]  & II\\
Current study& size & 1.3-6.0 & 22-91 & 0.6-2.5 & 1-420 & III\\
\end{tabular}
\end{ruledtabular}
\footnotetext[1]{For particles fed at the silo center, $W$ is half of the silo width.}
\footnotetext[2]{Flow rate is estimated as $Q=Q_m/(0.6\rho_m)$, where $Q_m$ is mass flow rate and $\rho_m$ is material density.}
\footnotetext[3]{Flow rate is estimated as $Q=F/0.6$, where $F$ is feed flow rate on net particle volume basis.}
\footnotetext[4]{$Q=Q_m/(0.6\rho_m)$, where $Q_m$ is mass flow rate and $\rho_m$ is estimated as 2.0 g/cm$^3$.}
\end{table*}
\end{center}

Although three final states of heap flow of bi-disperse granular materials (segregated, stratified, and mixed) have been observed and studied by different researchers, none of the past research appears to have investigated the dependence of the final particle distributions on a broad range of control parameters (see Table \ref{comparison}), including volumetric feed rate $Q$, silo width $W$, 2D silo gap thickness $T$, species size ratio $R=D_l/D_s$ ($D_l$ and $D_s$ are the large and small particle diameter, respectively), and absolute particle size. The focus of this research is to systemically explore how these parameters affect transitions between different final particle configurations.

Figure \ref{stages}(a) is a sketch of the top view of a three-dimensional (3D) silo, which typically comprises a vertical cylindrical container where granular material falls vertically along its centerline and flows radially in all directions down the heap, filling the container to its outer wall. Here, we use a quasi-two-dimensional (2D) silo, which can be thought of as a section of the 3D silo as shown in Fig.\ \ref{stages}(a). This makes it easy to observe the final particle distributions and minimizes the volume of particles needed. In the quasi-2D silo, rather than feed rate $Q$, the relevant flow rate is the 2D volumetric flow rate down the slope at the peak of the heap, defined as $q=Q/T$, which decreases linearly along the flow direction. As shown in Fig.\ \ref{stages}(b), filling of a silo proceeds in three stages. In stage I, a somewhat irregularly-shaped initial heap forms. Shortly, the heap becomes angled in stage II and grows laterally until it reaches the bounding end wall. In stage III, the laterally constrained heap rises steadily at a constant rise velocity $v_r$. We examine the steady filling stage (III), where the length of the flowing layer $L$ is constant.

In this Article, we present an experimental study of heap segregation of granular mixtures of binary \textit{spherical} particles differing only in size in a quasi-2D silo for a larger number and relatively wider range of experimental parameters than in previous work (Table \ref{comparison}). We observed all three final particle configurations - stratified, segregated and mixed. We find that for constant $R$, the transition from the stratified state to the unstratified state is controlled by $q$, while the transition from the segregated state to the mixed state is controlled by the rise velocity $v_r=q/W$. These transitions depend on size ratio $R$. For the first time, phase diagrams are plotted to illustrate the effects of $R$, $W$, and $q$ or $v_r$ on transitions between different final particle configurations. We further provide new insight for the occurrence of the stratified configuration for granular mixtures of \textit{spherical} particles compared to previous work \cite{Makse1997,Baxter} and propose a dimensionless velocity ratio - the ratio of rise velocity to particle percolation velocity - as a key control parameter for segregation.

In the remainder of this paper, Sec.\ \ref{setup} describes the experimental setup. Sections \ref{results} and \ref{discussion} present experimental results and discussion thereof. Section \ref{conclusion} presents our conclusions.

\section{Experimental setup}
\label{setup}

The quasi-2D silo in our experiments consists of a pair of 91 cm $\times$ 69 cm $\times$ 1.27 cm vertical rectangular plates: one is thick plate glass for observation and measurement purposes and the other is aluminum to reduce electrostatic charging. Vertical spacer bars were clamped between the parallel plates to control the silo gap thickness $T$ and to vary the silo width $W$. Granular mixtures were fed either at one end of the silo or at the centerline of the silo (the length of the flowing layer $L$ in the latter case is half of the former case). Similar results were observed independent of whether the feed was at one end of the silo (total silo width $W$) or at the centerline of the silo (total silo width 2$W$). For the results presented here, values for $W$ were 22 cm, 46 cm, 69 cm, and 91 cm. We note that the size of the silo at $W=91$ cm is comparable to small, full scale, industrial silos. The effect of the silo gap thickness $T$ on the transition between segregation and mixing is negligible, provided that $T$ is more than four large particle diameters. However, stratification depends sensitively on $T$, as discussed in the Appendix. Here, we use $T=$1.27 cm for all experiments, unless otherwise noted.

\begin{table}
\caption{\label{sizeratio} Binary mixtures of glass particles with equal mass fractions.}
\begin{ruledtabular}
\begin{tabular}{ccc}
Size ratio  & Large particles & Small particles \\
$R$ & $D_l$ (mm)\footnotemark[1] & $D_s$ (mm)\footnotemark[1] \\
\hline
1.3 & 2.00$\pm$0.08 & 1.51$\pm$0.09 \\
1.5 & 1.69$\pm$0.05 & 1.14$\pm$0.08 \\
2.0 & 2.00$\pm$0.08 & 1.00$\pm$0.06\\
2.2 & 1.10$\pm$0.06 & 0.50$\pm$0.05\\
3.4 & 1.69$\pm$0.05 & 0.50$\pm$0.05 \\
4.0 & 2.00$\pm$0.08 & 0.50$\pm$0.05\\
6.0 & 2.98$\pm$0.05 & 0.50$\pm$0.05 \\
\end{tabular}
\end{ruledtabular}
\footnotetext[1]{Mean particle diameter with standard deviation.}
\end{table}

Seven combinations of different-sized soda-lime glass particles with size ratios $R$ ranging from 1.3 to 6.0 were investigated (see Table \ref{sizeratio}). We limit $R$ to be less than $(2/\sqrt{3}-1)^{-1}= 6.464$, above which spontaneous percolation occurs \cite{Scott,savage,Lomine} (see Sec.\ \ref{limit}). To distinguish different species, different colors of particles were used. To ensure roundness and similar surface properties such as friction coefficient between different species, we purchased particles colored by the manufacturer (Sigmund Lindner GmbH, Germany). The granular mixtures in our experiments are composed of either metal-coated, surface-colored red and blue particles or clear and surface-colored black particles. Several trial experiments showed that final particle configurations are insensitive to the surface coatings of the two components. The material density of the particles is 2.59 g/cm$^3$.

A small auger feeder (Acrison Inc., NJ, USA) dispenses the granular mixtures into the silo. The auger feeder produces a stable and reproducible flow rate over a wide range of volumetric flow rates (1 to 420 cm$^3$/s). Flow rate is varied by controlling the rotation frequency of the motor and the diameter of the auger. The granular mixtures were composed of equal masses of large and small particles and were well-mixed upon filling the auger feeder. We performed several test experiments to measure the mass fraction of the two species of different-sized particles after discharge from the auger feeder at several different flow rates and size ratios. We found that the mass fraction of the two components remained at 50:50 (within $\pm$ 3$\%$) during the entire experiment. Before each experiment, anti-static spray (Sprayon, OH, USA) was applied to the glass wall to limit electrostatic effects. A digital camera in front of the glass wall recorded the filling and the final state of the heap. We performed more than 400 experimental runs, systematically varying control parameters (Table \ref{parameters}) including the feed volumetric flow rate $Q$, system size (silo width $W$ and silo thickness $T$), size ratio $R$, and absolute particle size.

\begin{table}
\caption{\label{parameters} Parameters}
\begin{ruledtabular}
\begin{tabular}{lll}
Parameter & Description & Type \\
\hline
$Q$ & 3D flow rate & controlled \\
$W$ & silo width & controlled \\
$T$ & silo gap thickness & controlled \\
$R=D_l/D_s$ & size ratio & controlled \\
$q=Q/T$ & 2D flow rate & calculated\\
$v_r=q/W$ & heap rise velocity & calculated\\
\end{tabular}
\end{ruledtabular}
\end{table}

\section{Results}
\label{results}

\begin{figure*}[]
\includegraphics[width=6in]{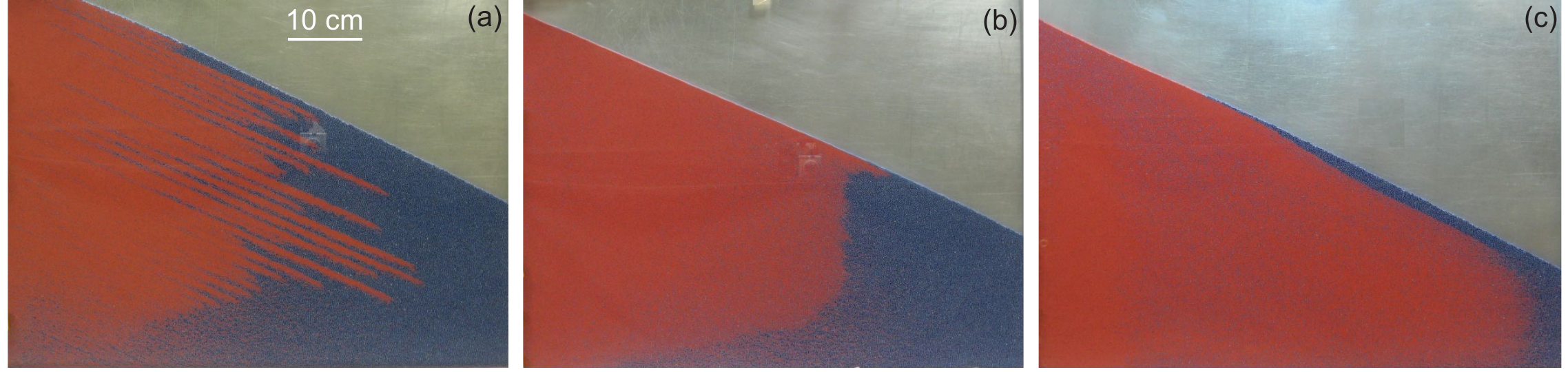}
\caption{\label{snapshot} (Color online) Three final particle configurations of different-sized granular mixtures in quasi-2D heap flow for $R=3.4$, $T$ = 1.27 cm, and $W$ = 69 cm (the free surface layer is associated with residual flow at the end of filling and should be ignored): (a) Stratification at $q$ = 0.8 cm$^2$/s; (b) Segregation at $q$ = 18.9 cm$^2$/s; (c) Near mixing at $q$ = 328 cm$^2$/s. Dark (blue online): 1.69 mm glass particles; light (red online): 0.5 mm glass particles. The width of each image is $W$.}
\end{figure*}

Stratified, segregated and mixed final states noted by previous researchers were all observed in our experiments for stage III, as shown in Fig.\ \ref{snapshot}. The thin free surface layer should be ignored as it is associated with residual flow at the end of filling. At small flow rates, similar but more pronounced stratification than reported in Baxter et al.\ \cite{Baxter} occurs [see Fig.\ \ref{snapshot}(a)]. The stratified state consists of alternating layers of small and large particles parallel to the flow direction, coexisting with segregation along the flow direction, where the downstream region of the heap contains mostly large particles and the upstream region of the heap close to the feed point contains mostly small particles. Stratification becomes weaker and eventually disappears as $q$ increases toward a critical value (to be discussed shortly). Above this critical value full segregation is observed, as shown in Fig.\ \ref{snapshot}(b). In this regime, the heap consists of two distinct regions: The downstream region contains of nearly all large particles, while the upstream region consists of a few large particles scattered in a sea of small particles. The boundary between these two segregated regions is narrow. As $q$ is further increased, the region of larger particles in the downstream portion of the heap shrinks and more large particles remain in the upstream portion of the heap. One might expect that at high enough $q$, the region containing all large particles disappears and a perfectly mixed state is achieved everywhere in the silo. However, due to limitations of our experimental apparatus, the silo cannot always be filled at a high enough flow rate $q$ to achieve perfect mixing. Instead, for most experiments at the highest achievable $q$, a near mixed state, see Fig.\ \ref{snapshot}(c), is obtained where only a narrow large particle region exists at the downstream end of the heap and the remainder of the heap is well-mixed.

\subsection{Stratification}
\label{stratification_transition}

\begin{figure}[]
\includegraphics[width=3.375in]{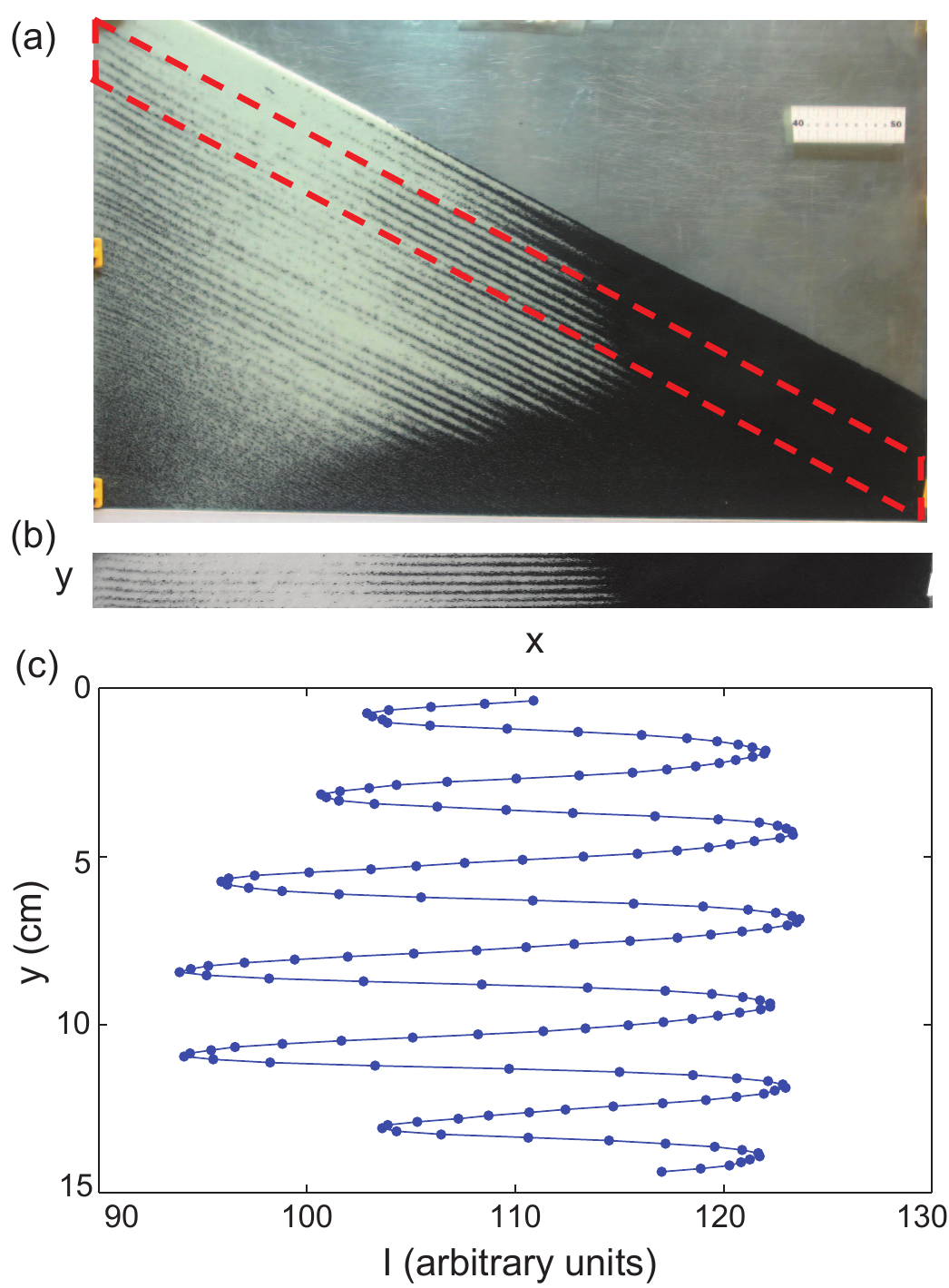}
\caption{\label{stratification} (Color online) Image processing method for quantifying stratification. (a) Image from the experiment with $W=$ 91 cm, $T=$ 1.27 cm and $q=$ 1.8 cm$^2$/s showing a stratified mixture of 0.5 mm (light) and 1.1 mm (black) glass particles. (b) Dashed parallelogram region from (a) transformed to a rectangular box. (c) intensity $I$ from (b) averaged over $x$ and plotted as a function of $y$.}
\end{figure}

Stratification [see Fig.\ \ref{snapshot}(a)] occurs at small flow rates over a wide range of size ratios and silo widths. We use the image intensity to quantify the final particle distributions for stratification for each experimental run. Since the particles used in the experiments have different colors, the local image intensity, $I$, is monotonically related to local particle concentration. In all our experiments, small particles have higher intensity than large particles, so higher local intensity implies higher local concentration of small particles.

Figure \ref{stratification} illustrates our method for measuring particle concentration for stratification. We study the region outlined in Fig.\ \ref{stratification}(a), which is located in stage III of the heap formation and excludes the free surface region associated with residual flow at the end of filling. The outlined region is transformed to a rectangular box [see Fig.\ \ref{stratification}(b)] by rotating by the angle of repose in the counter-clockwise direction and then ``shearing'' in the horizontal direction. The image intensity is averaged in the $x$-direction ($0 \le x \le L$) and plotted as a function of $y$ [Fig.\ \ref{stratification}(c)]. The intensity profile represents the variation of species concentration due to stratification; the periodic intensity oscillation in the $y$-direction corresponds to the alternating layers of small and large particles.

\begin{figure}[]
\includegraphics[width=3.375in]{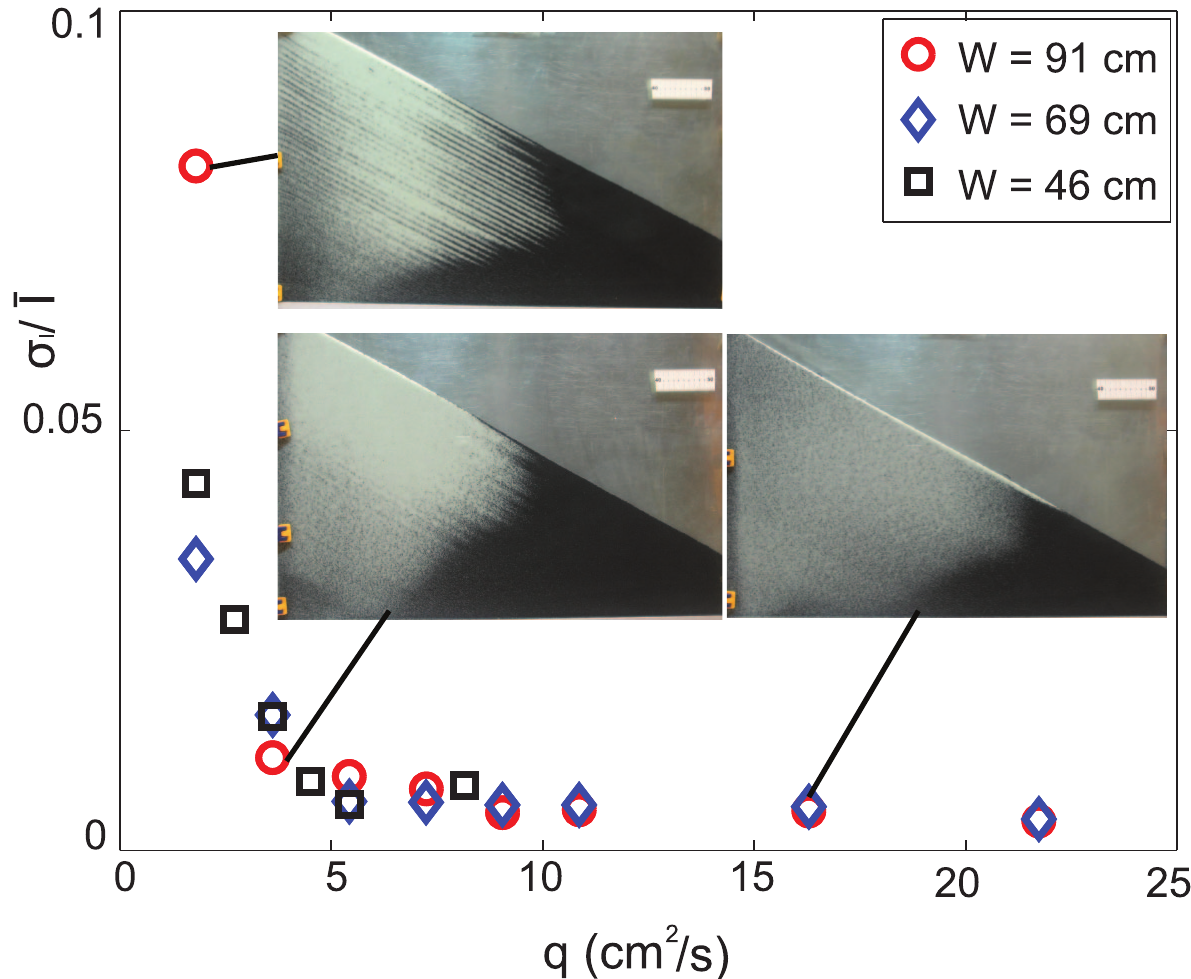}
\caption{\label{std_str} (Color online) $\sigma_I /\overline I$ showing decrease in stratification with increasing $q$ for $R=2.2$ at different $W$. Insets: images from experiments at the indicated data points, where $D_s=0.5$ mm for small light particles, $D_l=1.1$ mm for large black particles, and $W=91$ cm.}
\end{figure}

To further quantify the stratification globally and to determine the transition to segregation at different experimental conditions, we calculate the standard deviation of the intensity profile in Fig.\ \ref{stratification}(c) over the range of $y$ coordinates, $\sigma_I=\sqrt{\sum_{i=1}^N (I_i-\overline I)^2/(N-1)}$, where $\overline I$ is the mean intensity, $I_i$ is the intensity at row $i$, and $N$ is the number of pixel rows in the $y$-direction. Larger $\sigma_I /\overline I$ indicates a higher degree of stratification. When there is no stratification, $\sigma_I /\overline I$ goes to a constant residual value of 0.005 associated with variations in lighting intensity and random fluctuations of the layer concentration.

Figure \ref{std_str} shows $\sigma_I /\overline I$ as a function of $q$ for $R=2.2$. $\sigma_I /\overline I$ is significantly larger at small $q$ corresponding to strong stratification (long layers). As $q$ increases, $\sigma_I /\overline I$ decreases as the stratified layers become shorter and stratification weakens. $\sigma_I /\overline I$ decreases to a small constant value at a transitional 2D flow rate $q_t$, where stratification disappears and only segregation occurs. The transition from a stratified state to an unstratified state occurs around $q_t=$ 6 cm$^2$/s, independent of $W$. Similar trends are observed for all other $R$ except the smallest value considered, $R=1.3$, where no stratification occurs for all $q$ and $W$ tested.

\begin{figure}[]
\includegraphics[width=3.375in]{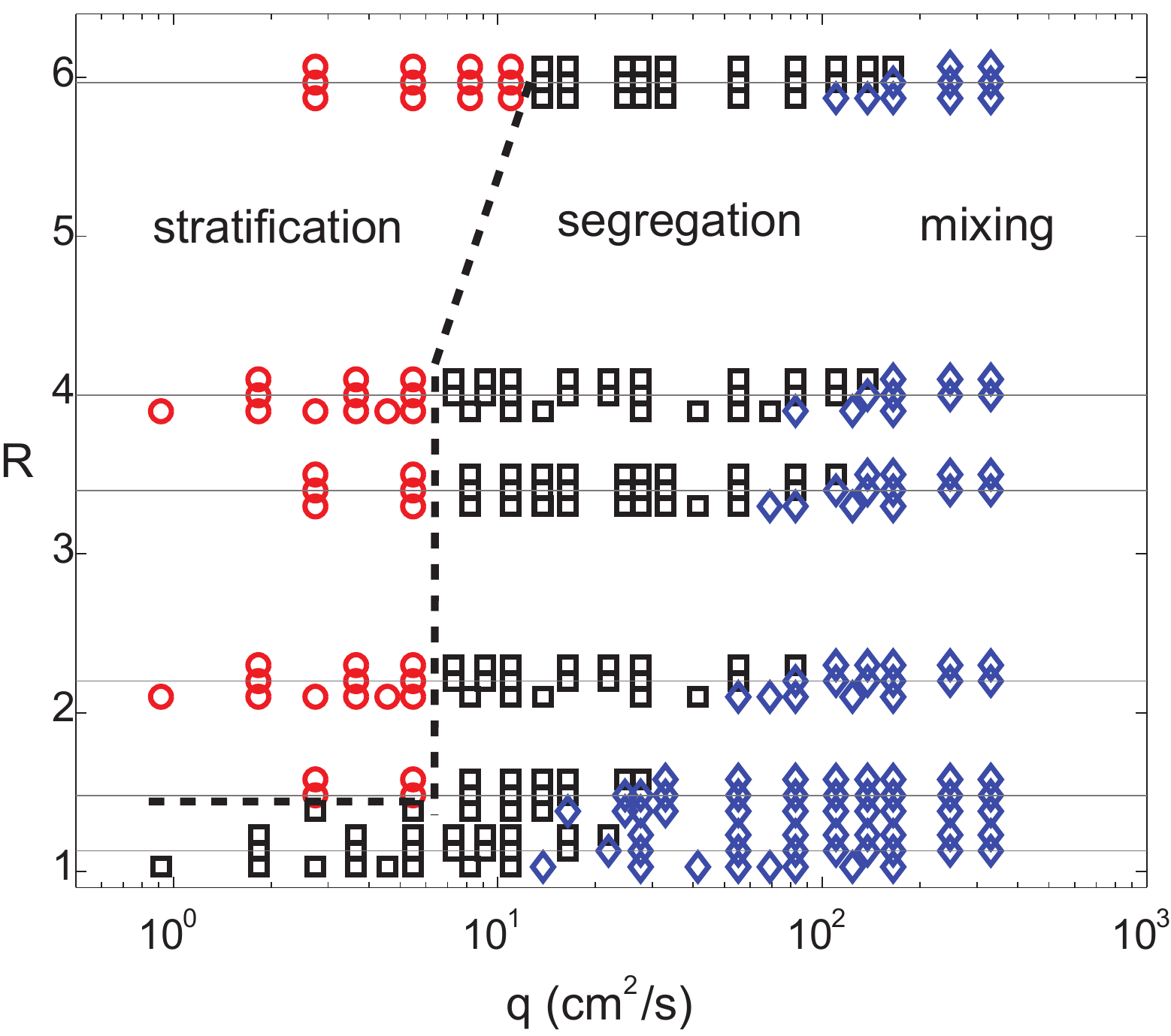}
\caption{\label{stra_r} (Color online) Phase diagram of final states (stratified: red $\bigcirc$; segregated: black $\Box$; mixed: blue $\Diamond$) in terms of $q$ and $R$ at three different $W$. Data are artificially offset in $R$ to show each data point for different $W$: from top to bottom, $W=$ 91 cm, 69 cm, and 46 cm, respectively. Horizontal solid lines denote the actual size ratio to guide the eye. Dashed lines segments mark the boundary between stratified and unstratified states. }
\end{figure}

The influence of $q$ , $W$, and $R$ on the transition from stratified to unstratified states is examined by plotting a phase diagram as a function of $q$ and $R$, shown in Fig.\ \ref{stra_r}. At each $R$, data for different values of $W$ are artificially offset in three rows representing from top to bottom, $W=$ 91 cm, $W=$ 69 cm, and $W=$ 46 cm, respectively, to show all of the data points. The three final states are distinguished by symbols and the dashed line indicates the boundary between the stratified state and the unstratified state. (We discuss the transition from segregation to mixing in Sec.\ \ref{segregation_transition}.)

For $2 \le R \le 4$, stratification occurs when $q$ is less than 6 cm$^2$/s. At these size ratios, the transition from the stratified state to the unstratified state occurs at the same transitional flow rate $q_t$ for all three values of $W$. For $R=6$, stratification is observed up to $q=$ 12 cm$^2$/s for all three values of $W$. This increase in $q_t$ is possibly because $R$ is close to the size ratio for spontaneous segregation, 6.464, so that other factors such as wall effects or horizontal segregation due to spanwise shear rate gradients \cite{Fan2011PRL} may significantly influence the stratification.

At small $R$, stratification diminishes. When $R=$1.3, no stratification occurs for all $W$. When $R=$ 1.5, stratification occurs only at $W=$ 69 cm and 91 cm; no stratification is observed at $W=$ 46 cm. Thus, $W$ may also affect the occurrence of the stratification at small $R$ as discussed further in Sec.\ \ref{discussion_stratification}.

\subsection{Segregation}
\label{segregation_transition}

\begin{figure}[]
\includegraphics[width=3.375in]{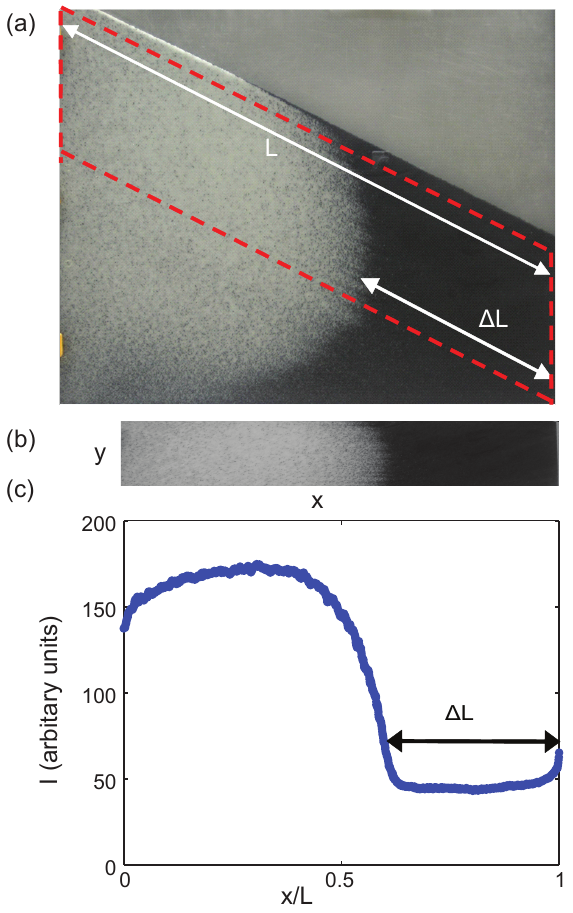}
\caption{\label{segregation} (Color online) Image processing method for quantifying segregation. (a) An image from the experiment with $W=$ 69 cm, $T=$ 1.27 cm, and $q=$ 10.9 cm$^2$/s for
a mixture of 0.5 mm (light) and 1.1 mm (black) glass particles. (b) The region in the dashed parallelogram from (a) transformed into a rectangle. (c) intensity $I$ from (b) averaged over $y$ and plotted as a function of $x/L$. $\Delta L$ denotes the width of the dark region of large particles at the end of the heap, and $L$ is the length of the flowing layer.}
\end{figure}

Full segregation occurs when $q$ increases beyond the transitional value $q_t$. A similar image processing method to that for stratification is used to quantify segregation. As shown in Fig.\ \ref{segregation}(a), the outlined region in stage III is considered. The parallelogram is transformed to a rectangle as in Fig.\ \ref{segregation}(b), and the image intensity is averaged in the $y$-direction and plotted as a function of $x$ as shown in Fig.\ \ref{segregation}(c) for a typical case. The upstream portion of the heap ($x/L \le 0.6$) has a smoothly varying concentration of particles of the two different sizes, whereas the downstream end ($x/L>0.6$) has a nearly uniform concentration of only large particles.

\begin{figure}[]
\includegraphics[width=3.375in]{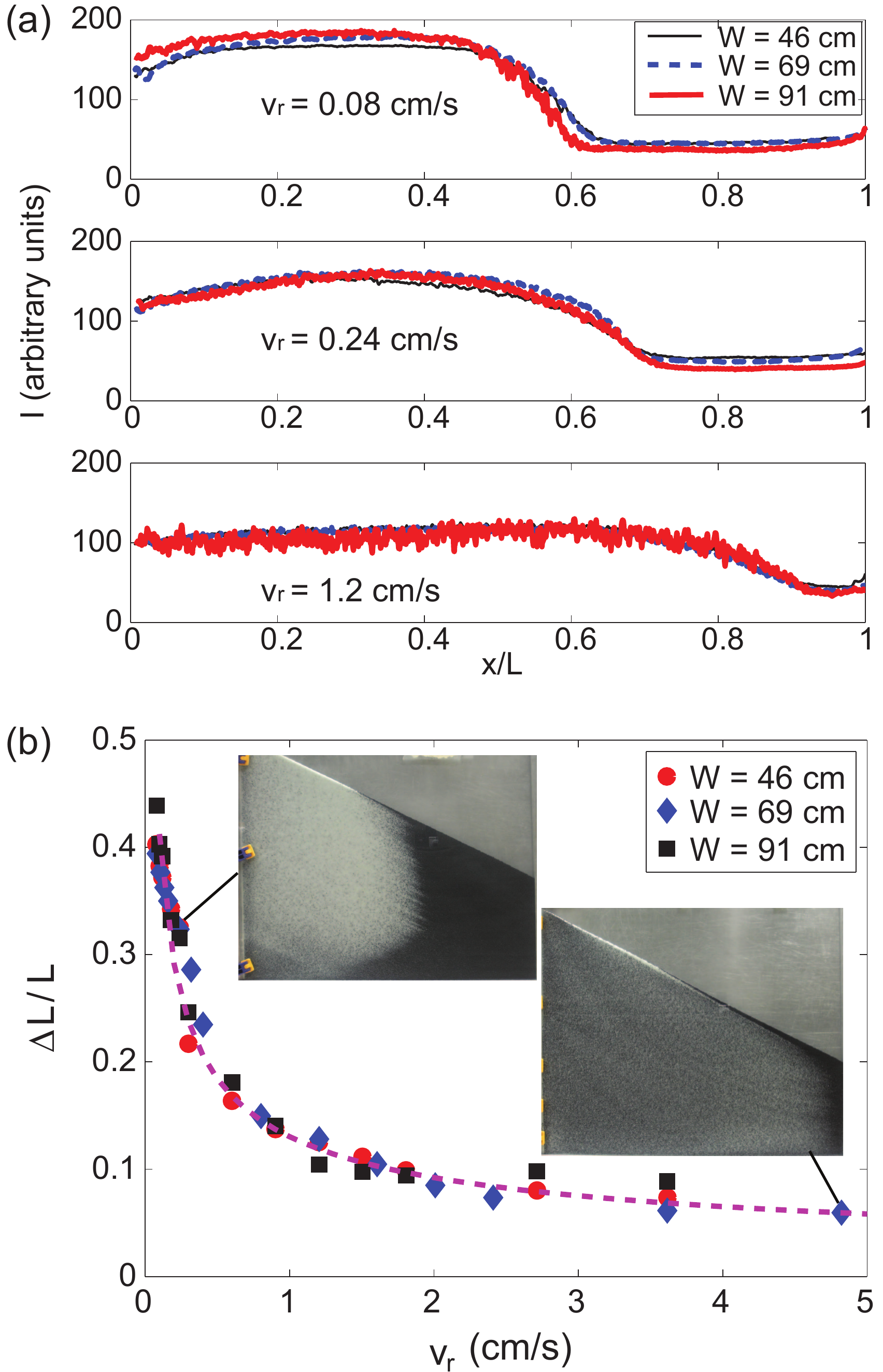}
\caption{\label{profiles_seg} (Color online)
(a) Intensity $I$ as a function of $x/L$ for a mixture of 0.5 mm (light) and 1.1 mm (black) particles at $v_r=$ 0.08 cm/s (top), 0.24 cm/s (middle) and 1.2 cm/s (bottom), respectively. (b) $\Delta L/L$ vs. $v_r$ for the same mixture at different silo widths $W$. Insets show images for $W=69$ cm and indicated data points. Dashed line is the fit $\Delta L/L = (v_r/a)^{-b}$, where $a=$0.017 cm/s and $b=$0.493.}
\end{figure}

Figure \ref{profiles_seg}(a) shows a series of intensity profiles plotted as a function of $x/L$ at different $W$ and $q$ at $R=2.2$. The profiles overlay one another for different values of $W$, as discussed shortly. Each profile represents the final state distribution of the two segregated species. The concentration of small particles is higher in the upstream region and the concentration of large particles is higher in the downstream region. The boundary between these two regions is narrow (less than 0.2$L$). Close to the feed zone of the heap ($x/L \le$ 0.2), the concentration of small particles is slightly smaller than in the rest of the upstream region [Such as Fig.\ \ref{segregation}(c) and Fig.\ \ref{profiles_seg}(a)]. This likely occur because the incoming particles discharged from the auger feeder start flowing from a nearly stationary state after falling onto the heap so that the segregation is weaker in this region.

As indicated in Fig.\ \ref{profiles_seg}(a), the similar intensity distributions in each plot correspond to identical rise velocities, $v_r$, but different silo widths $W$ and flow rate $q$. In stage III, $v_r=Q/(WT)=q/W$ and is independent of the angle of repose that changes slightly as $q$ or $R$ varies \cite{GDRMidi2004}. Thus, when $W$ and $q$ are varied together, $v_r$, instead of $q$, controls the final particle distributions for segregation. In other words, for a given mixture at the same $v_r$, small and large particles distribute similarly along the flow direction for different values of $W$ and $q$. When $v_r$ increases, more large particles stay in the upstream region so that the concentration of small particles decreases and the width of the downstream region of large particles $\Delta L$ decreases as well, as shown in Fig.\ \ref{profiles_seg}(a). However, the concentration of large particles in the downstream region of the heap does not change as $v_r$ increases (the intensity at large $x/L$ is constant at different $v_r$). Based on our experiments, there is almost always a region at the end of the heap close to the end walls containing nearly all large particles, regardless of the flow rate, silo width, and size ratio \footnote{There are two exceptions: (1) when $R<1.5$, the large particle region at the end of the heap is absent in many cases, which is discussed in Sec.\ \ref{limit}. (2) At large $D_s$ or small $W$, a narrow region of small particles appears at the end of heap due to another mechanism: the bouncing of particles. For all data presented in this paper, the bouncing-induced segregation is minimized by using small particles and large $W$ (see Sec.\ \ref{bounce_seg}).}.

To determine the transition from segregation to mixing as a function of $v_r$ and $R$, we quantify the degree of segregation using the normalized length of the large particle region $\Delta L/L$. This measure is similar to the approach of Shinohara et al.\ \cite{Shinohara1972}, who used $1-\Delta L/L$. In stage III, $\Delta L / L$ represents the approximate mass of segregated large particles relative to the mass of the mixture. Therefore, based on mass conservation, $0 \le \Delta L/L \le 0.5$, where $\Delta L/L=$ 0 means perfect mixing and $\Delta L/L=$ 0.5 means complete segregation.

\begin{figure}[]
\includegraphics[width=3.375in]{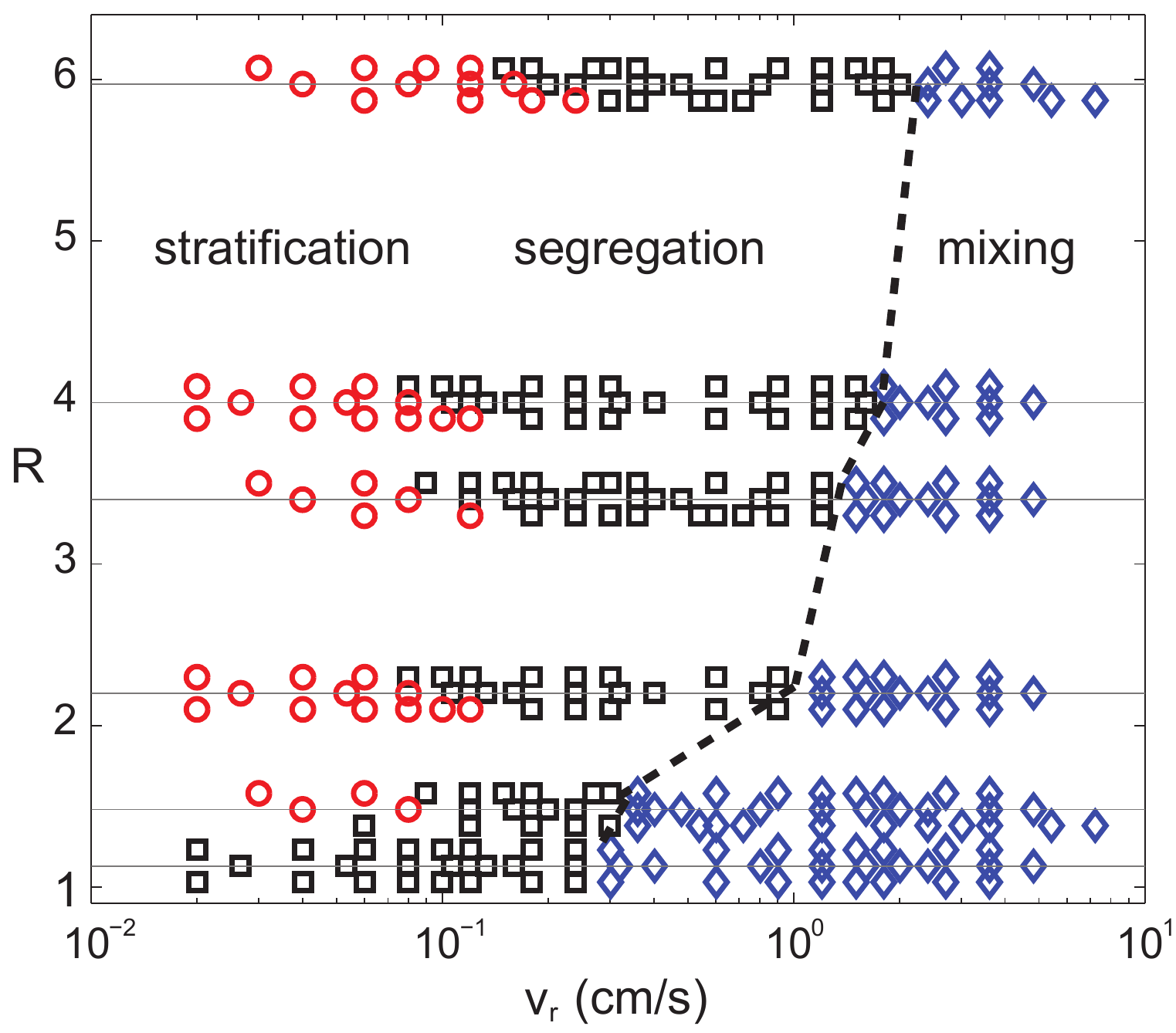}
\caption{\label{phase_diagram} (Color online) Phase diagram of final state (stratified: red $\bigcirc$; segregated: black $\Box$; mixed: blue $\Diamond$) in terms of rise velocity $v_r$ and size ratio $R$ at three silo widths $W$. Data are artificially offset in $R$ to show each data point for different $W$: from top to bottom, $W=$ 91 cm, 69 cm, and 46 cm, respectively. Horizontal solid lines denote the actual size ratio to guide the eye. Dashed line segments indicate the boundary between segregation and mixing.}
\end{figure}

Figure \ref{profiles_seg}(b) plots $\Delta L/L$ as a function of $v_r$ for different $W$ at $R=$ 2.2. At other size ratios similar trends occur. As $v_r$ is increased, $\Delta L/L$ decreases, indicating that segregation becomes weaker at higher $v_r$. For $v_r \ge 1.5$ cm/s, $\Delta L/L <0.1$, which means more than 80$\%$ of the large particles are mixed with small particles in the upstream region of the heap ($x/L < 0.9$). This corresponds to a mass ratio of small particles in this region of $56\%$. Curves of $\Delta L/L$ as a function of $v_r$ at different $W$ collapse onto a single curve, consistent with the collapse shown in Fig.\ \ref{profiles_seg}(a) for different $W$.

Due to limitations of our experimental apparatus (particles overflowing the entrance of the silo), a rise velocity greater than 10 cm/s (equivalent to filling the entire silo in less than 6 s) cannot be achieved. A perfect mixed state is therefor difficult to obtain at most $R$ except for $R=1.3$, where relatively good mixing is observed for $v_r >1$ cm/s. We fit the experimental data in Fig.\ \ref{profiles_seg}(b) to a power law as $\Delta L/L = (v_r/a)^{-b}$ using the least squares method for each size ratio [see Fig.\ \ref{profiles_seg}(b) for $R=2.2$], where the fitting parameters $a$ and $b$ depend on $R$. Based on this fit, a cutoff value for $\Delta L/L=0.15$ is selected to determine the transitional rise velocity between segregation and mixing, so that the boundary can be determined at different $R$. This value corresponds to a small particle mass ratio of $59\%$ in the region $x/L < 0.85$.

Using $\Delta L/L=0.15$ as the cutoff value between segregation and mixing, a phase diagram similar to that in Fig.\ \ref{stra_r}, but in terms of $v_r$ instead of $q$, is constructed to quantify the effects of $R$, $W$ and $v_r$ on the transition from segregation to mixing in Fig.\ \ref{phase_diagram}. The dashed line segments in the figure marks the approximate boundary between segregation and mixing. The phase diagram demonstrates that segregation transitions to mixing at the same $v_r$ for different $W$ at the same $R$. $v_r$ at the transition increases by roughly one order of magnitude as $R$ is increased from 1.3 to 6.0.

\section{Discussion}
\label{discussion}

As shown in Sec.\ \ref{results}, stratification generally occurs at low flow rates, and the transition from stratified to unstratified states depends on the flow rate $q$. In contrast, the transition from segregated to mixed states depends on the rise velocity of the heap $v_r$. In this section, we discuss possible mechanisms for these phenomena. Further, we briefly discuss, with respect to segregation, the limits of size ratio and the effect of particle bouncing after initial impact with the heap.

\subsection{Stratification dynamics}
\label{discussion_stratification}

As mentioned in Sec.\ \ref{Intro}, stratification has been most often observed in mixtures of large rough and small smooth particles as reported by Makse et al.\ \cite{Makse1997,MaksePRL}. The stratification is attributed to differences in the angle of repose: steeper for rough particles than for smooth spherical particles. Although Williams \cite{Williams1963} and Baxter et al.\ \cite{Baxter} showed some evidence of stratification for different-sized smooth spherical particles, the existence of stratification in mixtures of different-sized spherical particles has been debated.\footnote {Though Williams \cite{Williams1963} and Baxter et al.\ \cite{Baxter} observed stratification with smooth spherical particles, Makse et al.\ \cite{Makse1997,Cizeau} argued that some shape induced segregation may have occurred.} Furthermore, these works \cite{Williams1963,Baxter} only provide a few limited examples of stratification.

\begin{figure}[]
\includegraphics[width=3.375in]{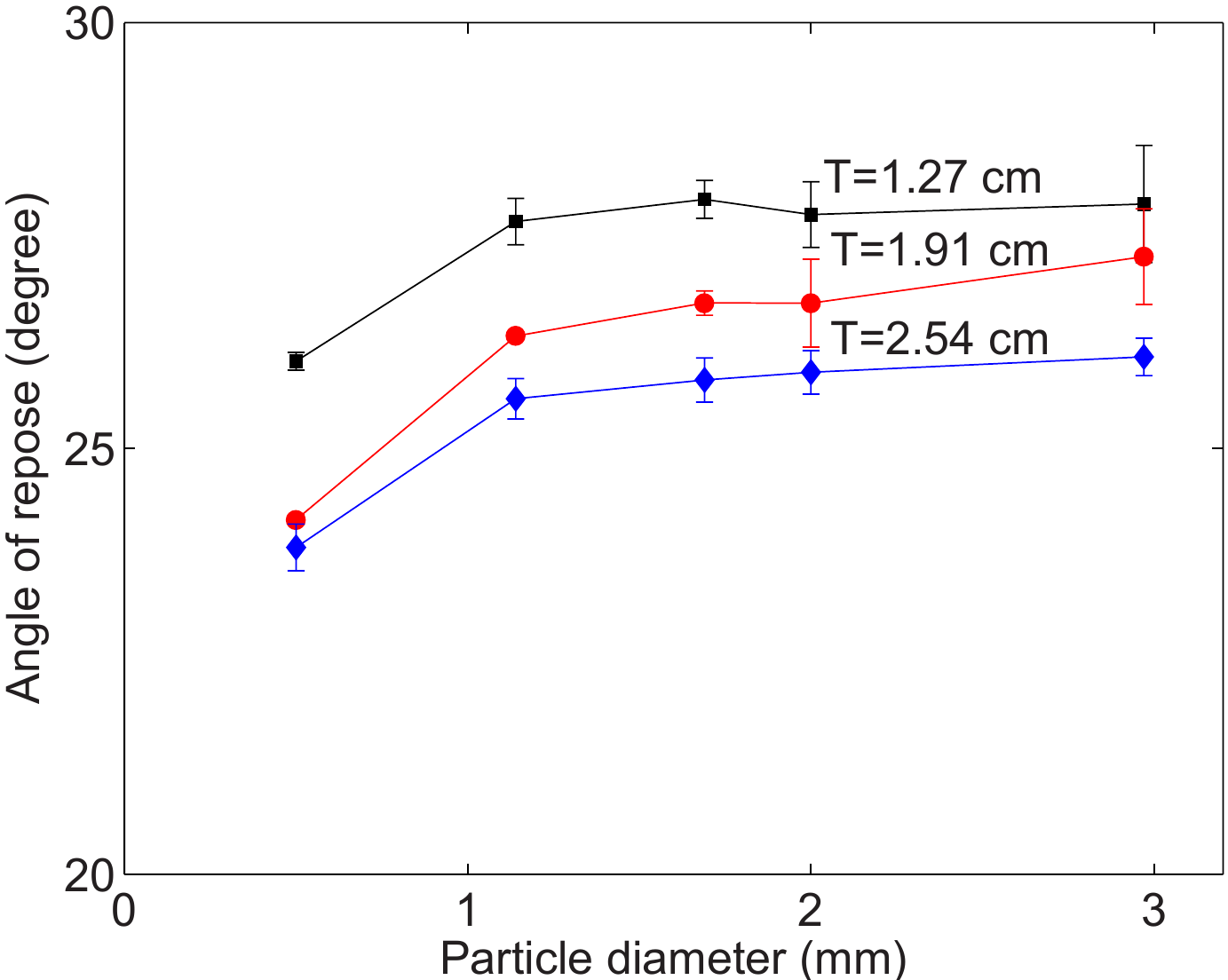}
\caption{\label{repose} (Color online) Static angle of repose for monodisperse glass particles at different silo gap thicknesses. Results are for 0.5 mm, 1.1 mm, 1.7 mm, 2 mm, and 3 mm monodisperse particles flowing into the silo at identical feed rate. The static angle of repose is measured after filling is stopped.}
\end{figure}

In this research, particles are all spherical so shape effects are excluded. Furthermore, except for the smallest particles, the angle of repose is nearly independent of particle size at the same silo gap thickness $T$, as shown in Fig.\ \ref{repose}. As described in Sec.\ \ref{stratification_transition}, stratification occurs for bi-disperse spherical particles over a wide range of flow rates, size ratios, and system sizes, which indicates a different mechanism for stratification than that proposed by Makse et al.\ \cite{Makse1997,MaksePRL}.

\begin{figure}[]
\includegraphics[width=3.375in]{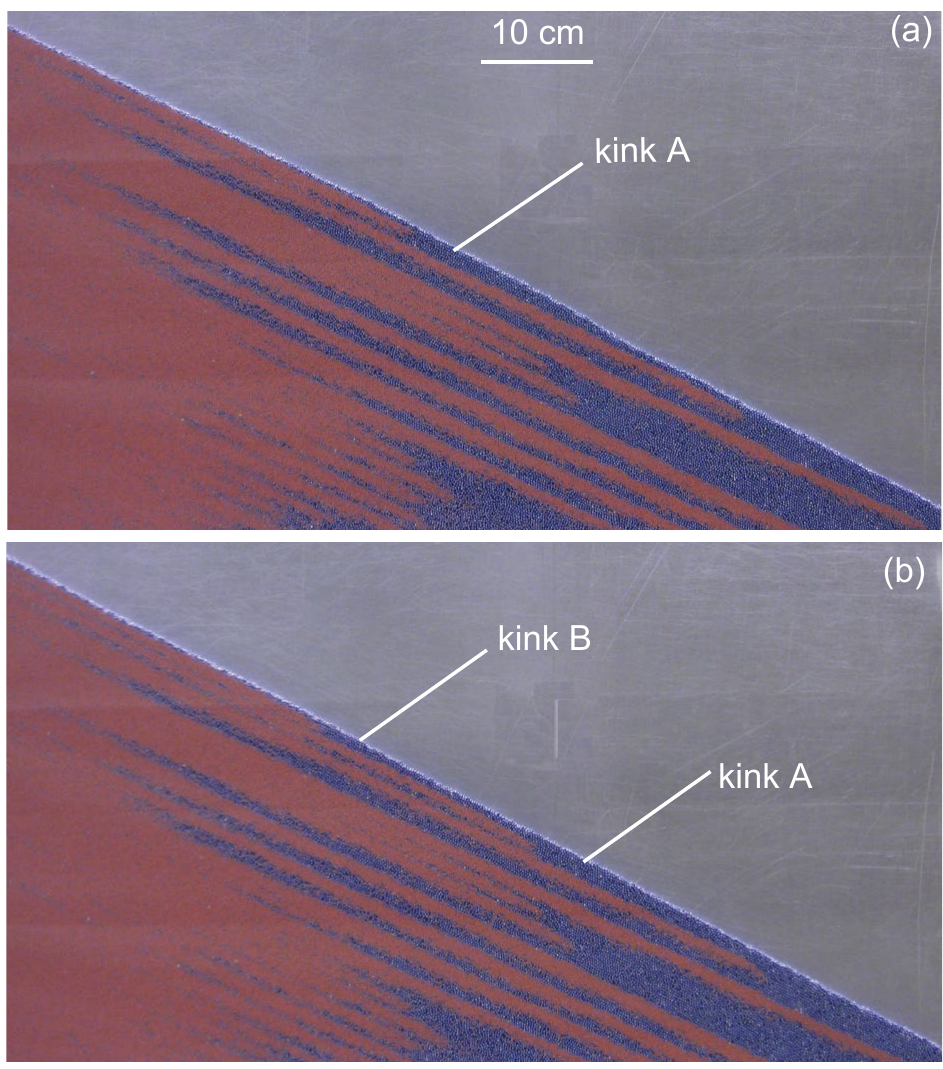}
\caption{\label{Str_snapshot} (Color online) Illustration of stratification formation at $R=3.4$, $W=91$ cm and $q=0.9$ cm$^2$/s. (a) and (b) show consecutive occurrences of two avalanches. Kink A is formed in the first avalanche, and it moves downstream when the second avalanche occurs, and during which kink B is formed. Dark (blue online): large particles; light (red online): small particles.}
\end{figure}

Careful observation of stratification for spherical particles in our experiments indicates that the appearance of layers of particles of different sizes is associated with the occurrence of discrete avalanches at low flow rates. As Fig.\ \ref{Str_snapshot} shows, during an avalanche, large (dark) particles segregate via percolation to the free surface and roll down the free surface forming the front of an avalanche. As time passes, the large particles in the front stop flowing and form a subtle kink in the slope of the heap [kink A in Fig.\ \ref{Str_snapshot}(a)]. Upstream from the kink, a stripe made of a thick layer of small particles at the bottom and a mono-layer of large particles at the surface forms [Fig.\ \ref{Str_snapshot}(a)]. Since the local angle of repose at the kink is larger than the static angle of repose, the kink is metastable. When a new avalanche occurs upstream of the kink and on top of the previous avalanche, kink (A) propagates further downstream, resulting in a part of the stripe associated with kink A reflowing with the arrival of the new avalanche [kink B in Fig.\ \ref{Str_snapshot}(b)]. As a result, the layer of small particles in the original stripe moves downward until the front of the avalanche, which consists of large particles, reaches the end wall of the silo, forming a layer of small particles above the larger particles from the previous kink (A) and below the larger particles from the current kink (B). In contrast, at high flow rates, surface flow is continuous, and large particles do not form kinks since there are no avalanches. Instead, large particles are continually advected to the downstream region of the heap. Small particles settle in the upstream region of the heap, resulting in the typical segregation pattern shown in Fig.\ \ref{snapshot}(b).

In addition, stratification is also limited by size ratio, silo width, and silo gap thickness. At small size ratios (e.g. $R=1.3$), segregation in the flowing layer evolves slowly that the initial stripe of small and large particles is weak. The silo also needs to be wide enough so that a kink can form somewhere along the slope of the heap. For example, no stratification occurs for $R=1.5$ and $W=46$ cm, while stratification does occur for large widths $W$ (see Fig.\ \ref{stra_r}). Silo gap thickness also affects the degree of stratification as discussed in the Appendix. A more detailed study of stratification of spherical particles is underway.

\subsection{Control parameters for segregation}
\label{discussion_seg}

The dynamics and mechanisms for segregation in heap flow have been previously studied \cite{Williams1963,Williams1968,Drahun,Shinohara1972,Shinohara1990} as discussed in Sec.\ \ref{Intro}. At its simplest, the competition between percolation of two different-sized components perpendicular to the flow direction and the advection of the mean flow in the flow direction determines the degree of final segregation. For instance, if percolation of small particles  downward through the flowing layer takes longer than the time to than the time to reach the downstream of the heap, more small particles will accumulate at the downstream region of the heap along with large particles resulting in a more mixed state. On the other hand, when small particles percolate quickly to the bottom of the flowing layer, they remain in the upstream region of the heap, resulting in stronger segregation.

\begin{figure}[]
\includegraphics[width=3.375in]{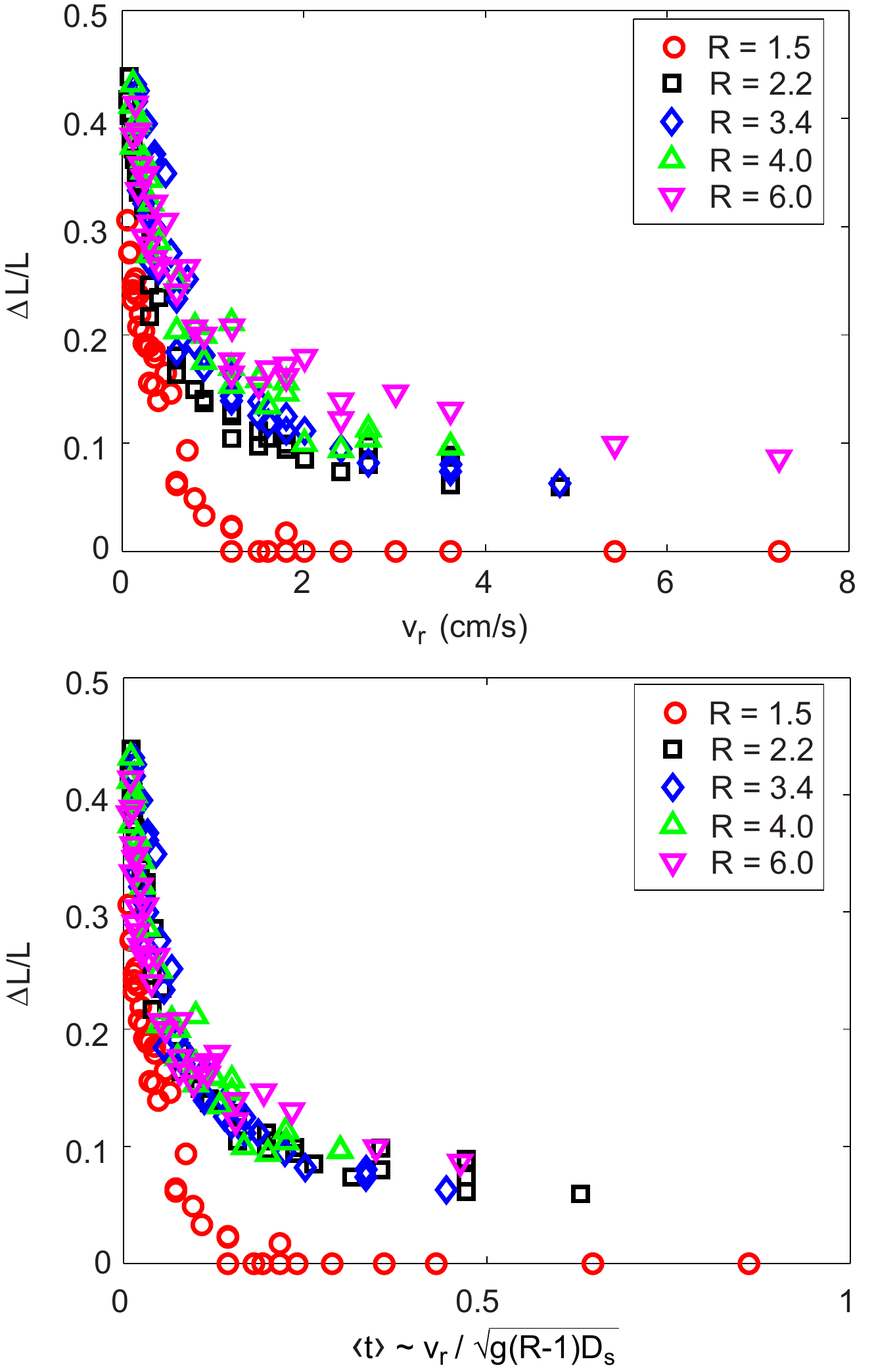}
\caption{\label{scale} (Color online) (a) $\Delta L/L$ vs. $v_r$ at different $R$ from 1.5 to 6. At each $R$, data for different values of $W$ use the same symbols. (b) $\Delta L/L$ as a function of the dimensionless time scale $\langle t \rangle \sim v_r/\sqrt{g(R-1)D_s}$.}
\end{figure}

Based on this picture, we define a dimensionless time $\langle t \rangle = t_{p}/t_{d}$, where $t_{p}$ represents the time scale for percolation, and $t_{d}$ represents the time scale for downstream convection by the mean flow. Based on this definition, $\langle t \rangle \gg 1$ indicates better mixing, while $\langle t \rangle \ll 1$ indicates stronger segregation. Assuming a constant average downstream velocity $\overline v_{d}$ and linear velocity profiles in the flowing layer for simplicity (similar to previous work on mono-disperse heap flow \cite{Boutreux,Khakhar2001}), we obtain $t_{d}=L/\overline v_{d}$, $t_{p}=\overline \delta /\overline v_{p}$ and $q=\overline v_{d}\overline \delta=v_{r}L$ (based on mass conservation in the flowing layer), where $\overline \delta$ is the mean thickness of the flowing layer and $\overline v_{p}$ is the mean percolation velocity. With these assumptions, the dimensionless time scale can be expressed as a velocity ratio

\begin{equation}
\label{segregation_time}
\langle t \rangle = v_{r}/\overline v_{p}.
\end{equation}

Equation (\ref{segregation_time}) shows that for a particular percolation velocity $\overline v_{p}$, which depends on the size ratio $R$, the degree of final segregation depends only on $v_r$. Increasing $v_r$ increases $\langle t \rangle$, indicating a decrease in segregation and an increase in mixing. For constant $v_{r}$, $\langle t \rangle$ decreases as $\overline v_{p}$ increases (e.g.\ $R$ increases). This means that at the same flow conditions, mixtures with large size differences should have a higher degree of segregation. Figure \ref{scale}(a) shows profiles of $\Delta L/L$ as a function of $v_r$ for several size ratios ($R=1.5,2.2,3.4,4,6$). As predicted by Eq.\ (\ref{segregation_time}), $\Delta L/L$, a measure of segregation, decreases as $v_r$ increases at constant $R$. Furthermore, at constant $v_r$, as $v_p$ is increased by increasing $R$, $\Delta L/L$ increases, indicating stronger segregation at larger $R$.


To further examine the dependence of segregation on $\langle t \rangle$ while varying $v_r$ and $\overline v_p$ simultaneously, a relation between the percolation velocity $\overline v_p$ and experimental parameters such as $R$ and strain rate at different $q$ is needed. Bridgwater and colleagues \cite{Cooke,Bridgwater2} systemically studied the influence of different parameters including the size ratio, density ratio, strain rate, and normal stress, on percolation velocities in various sheared systems. They found that particle size ratio has the greatest influence on the percolation velocity, while the others have little effect. However, an analytic relation is lacking. Here, to first order of approximation, we assume $\overline v_p$ depends only on $R$ along with gravity, which drives the flow. Figure \ref{scale}(b) shows $\Delta L/L$ as a function of $\langle t \rangle$ assuming $\overline v_p \propto \sqrt {g(R-1)D_s}$, where $R-1$ is used to assure zero percolation velocity for monodisperse particles. Curves of $\Delta L/L$ at different $R$ collapse well compared with Fig.\ \ref{scale}(a). However, for $R=1.5$, the scaling does not collapse the curves for $\langle t \rangle>0.07$ (corresponding to $v_r>0.5$ cm/s), presumably because in this situation, other mechanisms such as ordinary diffusion also play important roles in segregation. Thus, the expression we use for the percolation velocity is not always applicable.

\subsection{Upper and lower limits of size ratio}
\label{limit}
In this work we considered size ratios from 1.3 to 6.0, a relatively large range compared to previous studies (Table \ref{comparison}) of bidiperse mixtures of spherical particles. We choose $R \le 6$ for two reasons. First, as shown in Refs.\ \cite{Scott,savage,Lomine} and references therein, when $R$ is larger than $(2/\sqrt{3}-1)^{-1}= 6.464$, a small particle can percolate through the smallest voids between three large particles without external agitation. Thus, the mechanism for segregation is different from that at the smaller $R$ investigated here. Further, as mentioned by Fan and Hill \cite{Fan2011PRL}, a shear gradient can also drive segregation in the spanwise direction across the silo. At larger $R$ , the shear effect is more important than at smaller $R$. During the experiments, at $R=6$, we observed signs of both effects. Small particles preferentially fill the voids between large particles at the side walls, which causes a variation of particle concentration between the side walls. Based on observations at the top of the free surface, large particles also tended to gather toward the middle of the gap away from the side walls for $R=6$, probably due to the shear effect \cite{Fan2011PRL}.

\begin{figure}[]
\includegraphics[width=3.375in]{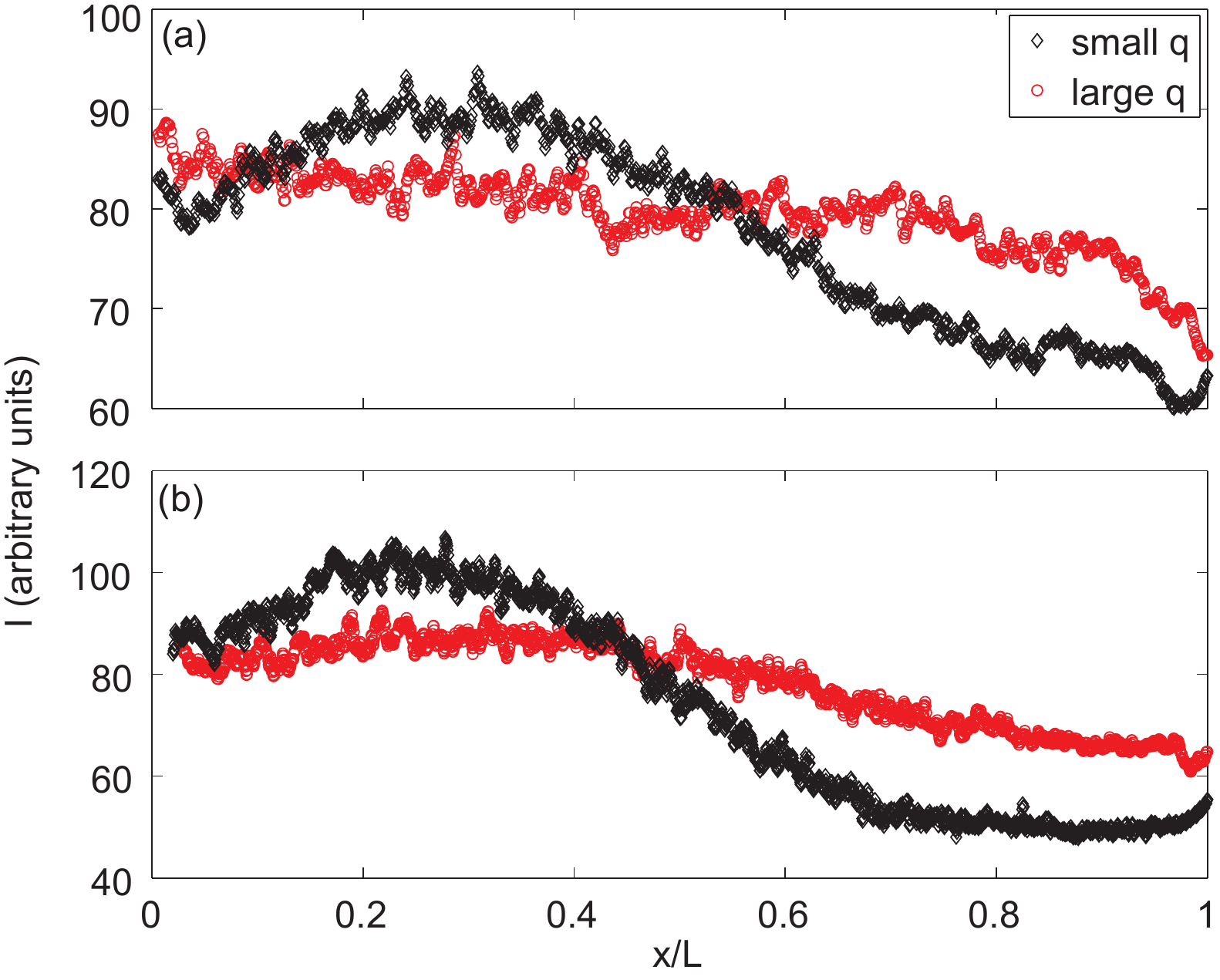}
\caption{\label{R133} (Color online) Intensity as a function of $x/L$ for $R=1.3$ at $q=1.8$ cm$^2$/s (black $\Diamond$) and $q=21.6$ cm$^2$/s (red $\bigcirc$) for (a) $W=46$ cm and (b) $W=91$ cm.}
\end{figure}

At small size ratios (e.g.\ $R=$ 1.3), no stratification is observed because of the small size difference, as discussed in Sec.\ \ref{discussion_stratification}. Segregation also becomes much weaker at small size ratios and exhibits different characteristics than at other size ratios. As mentioned by Goyal and Tomassone \cite{Goyal}, when $R$ is smaller than a critical value of 1.4, the concentration of each species changes continuously along the heap, as shown for small $q$ in Fig.\ \ref{R133}(a) for $R=1.3$. For $R>1.4$, segregation is more complete and boundaries between the two segregated regions are sharper, as shown in Fig.\ \ref{profiles_seg}. Our results show that the transition between these two different segregated states additionally depends on $q$ and $W$, as shown in Fig.\ \ref{R133}. At $W=46$ cm for $R=1.3$ and within the entire range of $q$ in our experiments, continuously varying segregation is always observed [Fig.\ \ref{R133}(a)]. However, for $W=91$ cm, more complete segregation occurs at lower $q$ and continuously varying segregation occurs at higher $q$ [Fig.\ \ref{R133}(b)]. One possible reason for this behavior is that at small $R$, ordinary diffusion, which causes re-mixing of different components, may become comparable in importance to percolation over shorter $W$. Diffusion in flowing granular materials is related to the flow kinematics such as local shear rates \cite{Natarajan}, which are closely associated with the flow rate and system size. At small flow rates or large silo widths, diffusion could be weaker than percolation so that segregation is stronger. Further study of the segregation mechanism at small size ratios $R<1.5$ is currently underway.


\subsection{Bouncing-induced segregation}
\label{bounce_seg}
\begin{figure}[]
\includegraphics[width=3.375in]{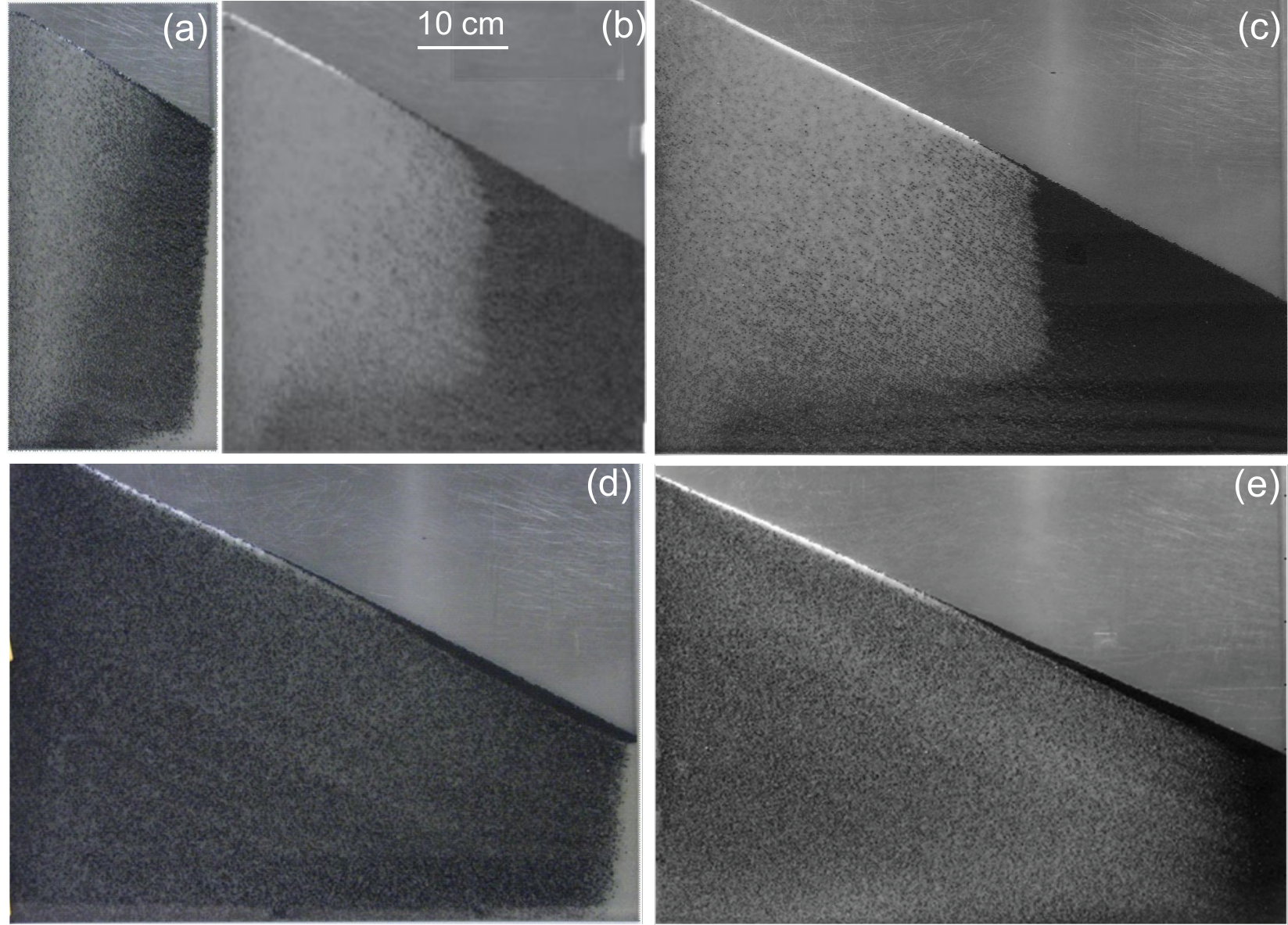}
\caption{\label{bounce} (a)-(c) Images showing reduction in bouncing-induced segregation with increasing silo width $W$ at $R=4$ with 2 mm (black) and 0.5 mm (light) glass particles and $q=14$ cm$^2$/s. (a) $W=$ 22 cm, (b) $W=$ 46 cm, and (c) $W=$ 69 cm. (d) and (e) Images at similar size ratio ($R \approx 2$) showing reduction in bouncing-induced segregation with decreasing absolute particle size at $W=69$ cm and $q=140$ cm$^2$/s. (d) 2.0 mm (black) and 1.0 mm (light) particles, (e) 1.1 mm (black) and 0.5 mm (light) particles.}
\end{figure}

When the falling particles feeding the heap impact the top of the heap, they sometimes bounce away from the top of the heap toward the end of the silo. Since small particles generally gain momentum when colliding with large particles (similar to when a lighter golf ball collides with a heavier basketball), they tend to bounce further down the heap than large particles. This can result in segregation that is reversed and unrelated to that in the flowing layer. Bouncing-induced segregation, briefly mentioned by Drahun and Bridgwater \cite{Drahun}, causes small particles to segregate to the downstream region of the heap, which is, of course, opposite to what occurs during the free surface segregation studied in this paper.

Several factors can influence bouncing-induced segregation including the silo width, the fall height of the particle feed stream, and the relative and absolute particle size. Figure \ref{bounce} shows the influence of these factors on bouncing-induced segregation. In Figs.\ \ref{bounce}(a)-(c), $W$ is varied while all other parameters are fixed. For the smallest silo width, $W=$ 22 cm [Fig.\ \ref{bounce}(a)], a narrow vertical band of small light-colored particles forms at the downstream end of the heap (right wall). This is opposite to what is observed under normal free surface segregation conditions. When $W$ is increased to 46 cm and 69 cm [Figs.\ \ref{bounce}(b)-(c)], no small particles are observed adjacent to the right end wall. Even though bouncing of particles still occurs at larger $W$, the bouncing particles re-enter the flowing layer before reaching the end of the silo, so that bouncing-induced segregation is negligible. The effect of fall height on bounce-induced segregation can be seen by considering a single experimental run [see Fig.\ \ref{bounce}(a)]. As the heap grows, the fall height decreases, so that the bouncing-induced segregation becomes weaker. Consequently, in Fig.\ \ref{bounce}(a), the small particle region at the right wall of the silo becomes narrower from bottom to top. At the smallest fall height near the end of the experiment and corresponding to the top of the vertical band of particles, there are only large particles close to the end wall, indicating a strong decrease in bouncing-induced segregation. In Figs.\ \ref{bounce}(b)-(c), in addition to the fact that no small particles are observed in the downstream regions of the silo next to the right wall, the boundaries between large and small particle region are vertical (essentially independent of the fall height), indicating that bouncing-induced segregation has negligible influence on the final particle distributions.

Figures \ref{bounce}(d)-(e) compare segregation when the granular mixtures have similar size ratios ($R \approx 2$) but different absolute sizes [2 mm and 1 mm in Fig.\ \ref{bounce}(d) and 1.1 mm and 0.5 mm in Fig.\ \ref{bounce}(e)]. The mixture with the larger absolute size exhibits stronger bouncing-induced segregation [small particles accumulate at the downstream wall of the silo in Fig.\ \ref{bounce}(d)] than the mixture of smaller absolute size [no small particles accumulate at the downstream wall of the silo in Fig.\ \ref{bounce}(e)]. The inertial and gravitational forces of the particles are proportional to the cube of particle radius, while the air drag is at most proportional to the square of particle radius (at high Reynolds numbers) \cite{Batchelor}. Therefore, the ratio of inertial or gravitational forces to air resistance is greater for the mixture of larger particles than for the mixture of smaller particles. As a result, the small particles in the mixture bounce farther after impact [Fig.\ \ref{bounce}(d)].

This cursory study of bouncing-induced segregation demonstrates that bouncing-induced segregation can compete with free surface segregation and may result in different final segregation patterns. However, as long as the silo width is large enough, bouncing-induced segregation is minimized and free surface segregation dominates the final particle distributions. To further study the effects of bouncing-induced segregation in heap flow, alternative methods to feed particles to the top of the heap (e.g. similar to that in Ref.\ \cite{Samadani}) may be needed to isolate the effects of these two different segregation mechanisms..

\section{Conclusions}
\label{conclusion}

In this paper, we have shown that three different final configurations in heap flow - stratified, segregated and mixed - can be obtained by controlling flow properties, either the flow rate $q$ or the rise velocity of the heap $v_r$. Stratification is associated with discrete avalanches at low flow rates ($q$ smaller than $\sim$10 cm$^2$/s for spherical glass particle mixtures) at most size ratios ($R>1.4$). The silo width $W$, or alternatively the flowing layer length $L$, should also be large enough that a kink can form along the slope of the heap (instead of at the end of the heap) during the avalanche. The transition from a stratified to an unstratified state is governed by $q$. When $q$ is larger than a transitional value dependent on $R$, the heap flow transitions from a discrete avalanche regime to a continuous flow regime with segregation in which neither kinks nor stratification are observed. The degree of segregation is determined by competition between advection by the mean flow and percolation through the flowing layer. This competition can be characterized by the ratio of the rise velocity of the heap $v_r$ to the percolation velocity $\overline v_p$, which mainly depends on $R$. At the same $R$, as $v_r$ increases, the degree of segregation decreases and eventually transitions to a mixed state. The transitional rise velocity becomes larger as $R$ increases.

There are three major points, among others, that this study raises: (i) Stratification of different-sized spherical particles is observed for a wide range of flow rates and size ratios, but the dynamics of stratification appear different from those for stratification of different size and shape particles observed in Makse et al.\ \cite{Makse1997,MaksePRL}. These apparent differences in the physical mechanisms for stratification need further study; (ii) This study shows that other mechanisms play important roles in heap segregation for the following situations: a) At small size ratios ($R<$1.5), ordinary diffusion appears to become comparable to percolation, suggesting that a different segregation configuration occurs (continuously varying segregation). b) For $R$ near the large size ratios at which spontaneous percolation can occur ($R \approx 6.5$), wall effects or shear-induced segregation may cause spanwise segregation between the side walls, which requires study of segregation/stratification in the bulk of a fully 3D silo. (iii) When particles have sufficient impact velocity or the silo is not wide enough, bouncing of particles after heap impact can cause segregation opposite that of the usual segregation in the flowing layer.

We have only studied quasi-2D configurations of heap flow. While we expect that these results are indicative of those in 3D heaps, further work is needed to confirm this. Furthermore, we have only considered the steady filling stage of heap formation [see Fig.\ \ref{stages}(b)], leaving initial heap formation and heap growth for future investigation.


\begin{acknowledgments}
We are grateful for the laboratory assistance of Emre Yildiz and helpful discussions with Karl Jacob and Ben Freireich. We also acknowledge the financial support of The Dow Chemical Company.
\end{acknowledgments}

\appendix*\section{Effects of quasi-2D silo gap thickness $T$}

\begin{figure}[]
\includegraphics[width=3.375in]{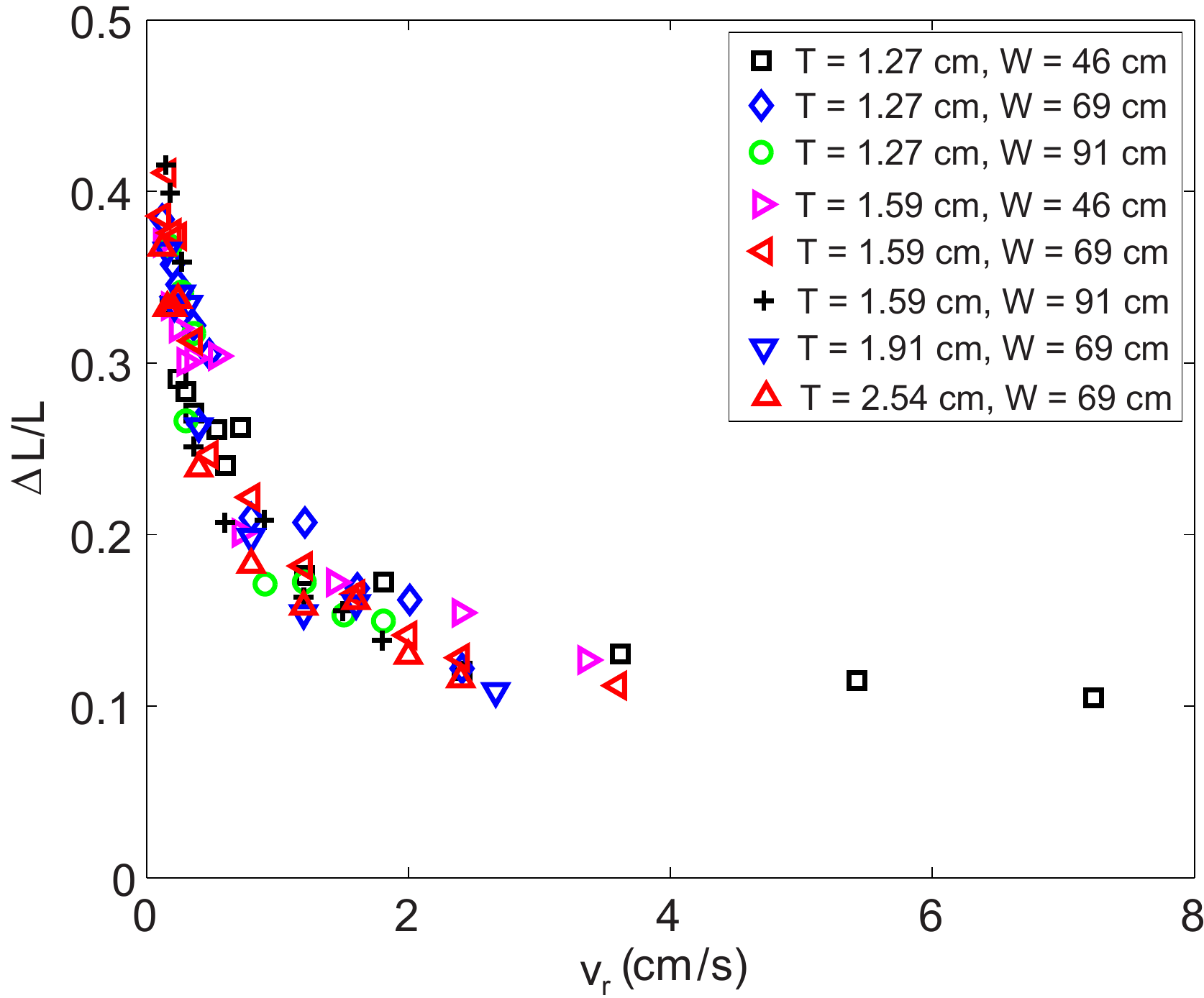}
\caption{\label{thickness_effect} (Color online) $\Delta L/L$ vs.\ $v_r$ for mixtures of 2.98 mm and 0.5 mm particles ($R=6$) at different silo gap thicknesses $T$ and silo widths $W$, showing segregation is insensitive to $T$ for $T>4D_l$. }
\end{figure}
The effect of quasi-2D silo gap thickness $T$ on stratification and segregation was considered for   several binary mixtures at various $q$ to assure that $T$ has no significant influence on the results. Silo thickness $T$ was varied from 0.64 cm to 2.54 cm (corresponding to $3 < T/D_l < 12$) and $q$ was varied from 1.2 cm$^2$/s to 331 cm$^2$/s. Measurements of stratification and segregation are carried out using the same criteria as those in Sec.\ \ref{results}.

The effects of $T$ on segregation are shown in Fig.\ \ref{thickness_effect} by plotting $\Delta L/L$ vs. $v_r$ for $R=6$. Within the range of $T$ investigated, the degree of segregation does not depend on $T$, as long as $T > 4D_l$. The same trends are observed for other values of $R$. In contrast, stratification is influenced by $T$, but no clear trend is observed. For example at $R=6$, when $T$ increases from 1.27 cm to 2.54 cm, $\sigma/\overline I$ decreases at the same $q$. At $T=2.54$ cm, $\sigma/\overline I$ is fairly small over the entire range of $q$, indicating that at this $T$, stratification is barely observable for all experimental parameters. However, at $R=2$ or $R=3.4$, and certain values of $q$, $\sigma/\overline I$ at larger $T$ is larger than at smaller $T$, indicating that stratification does not always monotonically decrease when $T$ increases at all $R$. The mechanism for the dependence of stratification on $T$ remains unclear and needs further investigation. In this paper, we use $T=1.27$ cm for all experiments as it is sufficient for achieving $T$-independent segregation and it minimizes the volume of particles needed to perform the experiments.

%

%

\end{document}